# A protonated brownmillerite electrolyte for superior low-temperature proton conductivity

Nianpeng Lu[1†], Yujia Wang[1†], Shuang Qiao[1†], Hao-Bo Li[1†], Qing He[2], Zhuolu Li[1], Meng Wang[1], Jingzhao Zhang[3], Sze Chun Tsang[3], Jingwen Guo[1], Shuzhen Yang[1], Jianbing Zhang[1], Ke Deng[1], Ding Zhang[1, 4], Jing Ma[5], Yang Wu[1, 6], Junyi Zhu[3], Yoshinori Tokura[7], Ce-Wen Nan[5], Jian Wu[1, 4*] and Pu Yu[1, 4, 7*]

[1]State Key Laboratory of Low Dimensional Quantum Physics and Department of Physics, Tsinghua University, Beijing, 100084, China

[2]Department of Physics, Durham University, Durham DH1 3LE, United Kingdom

[3]Department of Physics, The Chinese University of Hong Kong, Shatin, New Territories, Hong Kong, China

[4]Collaborative Innovation Center of Quantum Matter, Beijing 100084, China

[5]State Key Lab of New Ceramics and Fine Processing, School of Materials Science and Engineering, Tsinghua University, Beijing 100084, China

[6]Tsinghua-Foxconn Nanotechnology Research Center, Tsinghua University, Beijing 100084, China

[7]RIKEN Center for Emergent Matter Science (CEMS), Wako 351-198, Japan

†These authors contributed equally to this work.

**Design novel solid oxide electrolyte with enhanced ionic conductivity forms one of the Holy Grails in the field of materials science due to its great potential for wide range of energy applications[1-5]. Conventional solid oxide electrolyte typically requires elevated temperature to activate the ionic transportation, while it has been increasing research interests to reduce the operating temperature due to the associated scientific and technological importance[6-14]. Here, we report a conceptually new solid oxide electrolyte, $HSrCoO_{2.5}$, which shows an exceptional enhanced proton conductivity at low temperature region (from room temperature to 140 °C). Combining both the experimental results and corresponding first-principles calculations, we attribute these intriguing properties to the extremely-high proton concentration as well as the well-ordered oxygen vacancy channels inherited from the novel crystalline structure of $HSrCoO_{2.5}$. This result provides a new strategy to design novel solid oxide electrolyte with excellent proton conductivity for wide ranges of energy-related applications.**

Solid oxide proton conductor holds significant promise as the electrolyte because of its enhanced ionic conductivity at reduced operating temperature [8, 9]. In conventional solid oxide proton conductors (e.g. perovskite $BaZr_{0.8}Y_{0.2}O_{3-\delta}$) [11, 12], the existence of oxygen vacancy is the prerequisite condition to incorporate hydrogen into the form of hydroxide anion (the left panel of **Fig. 1a**), in which the vibration of H-O bond activate the proton transportation through the hopping mechanism at elevated temperature. However, the proton content in this case is strongly limited by the concentration of oxygen vacancy, which is typically in the order of a few percent. Recently, a new type of proton conductor was reported in the form of protonated perovskite nickelate ($HSmNiO_3$, or HSNO)[14], whose crystal lattice can host a great number of protons without the requirement of oxygen vacancy, as shown in the right panel of **Fig. 1a**. However, the proton conductivity of this compound is limited to a notable value of ~0.01 S·cm$^{-1}$ at the intermediate temperature region (~500 $^{o}$C), probably due to the insufficiency of vacant sites to assist the ionic hopping [15, 16]. Hence, the main challenge still remains, and inspires us to seek for novel mechanisms and material model systems with superior ionic conductivity at lower temperature (in particular around room temperature) region.

Here, we report a conceptually new solid oxide electrolyte, protonated brownmillerite strontium cobaltite (i.e., $HSrCoO_{2.5}$ (HSCO)), as shown in **Fig. 1b**. The significance of such material is its intrinsically ordered vacancy channels [17, 18], which provide an optimal condition for the intercalation and diffusion of protons within the crystal structure. By successfully overcoming the limitations of both proton concentration and vacant site, HSCO exhibits a dramatically improved proton conductivity with small ionic activation energy (0.27 eV) around room temperature region.

In this study, HSCO thin films with different thicknesses were obtained through protonating the as-grown $SrCoO_{2.5}$ (SCO) thin films using a recently demonstrated ionic liquid gating (ILG) method[18]. With confirmed high-quality protonated HSCO samples (**Supplementary Figs. 1** and **2**), we carried out impendence spectroscopy measurements to investigate its intrinsic proton conductivity (**Method Section** and **Supplementary Fig. 3**). The measurements were carried out with temperature up to 140 °C, due to the fact that HSCO would transform back to SCO with temperature higher than 160 °C (**Supplementary Fig. 4**). The Nyquist plots of the impedance data (**Fig. 1c**) clearly exhibit a linear relationship between the real and imaginary parts with the slope of ~1 at low frequency region, which indicates the polarization of electrode [19] on account of the protonic diffusion at the triple-phase boundary around the anode. At high frequency region, the plots develop into semi-circles mainly due to the proton conductivity of HSCO. By fitting the impendence spectra with the equivalent circuit shown in **Supplementary Fig. 5**, the temperature dependent proton conductivity of HSCO was exacted and plotted with the ionic conductivities of other electrolytes in **Fig. 1d**. Excitingly, the observed proton conductivity in HSCO at low temperature region (room temperature to 140 °C) is much larger than that of HSmNiO$_3$[14] and other solid oxide electrolytes at even much higher temperature region [20-23] (**Fig. 1d**).

Moreover, the proton conductivity ($\sigma$) in HSCO is dramatically enhanced with the increasing of temperature and nicely follows the Arrhenius relationship, $\sigma = \sigma_0 \exp(\frac{-E_a}{k_B T})$, where $k_B$ is the Boltzmann constant, and $T$ is the absolute temperature. Accordingly, the activation energy ($E_a$) is estimated as 0.27 eV, comparable with that of HSNO (~ 0.3 eV) [14] and other solid acid protonic conductors [20]. More importantly, the low-temperature ionic conductivity of HSCO, being 0.028 S·cm$^{-1}$ and 0.33 S·cm$^{-1}$ at 40 °C and 140 °C respectively, is even competitive with that of the widely applied proton exchange membranes at similar temperature [4, 15, 24, 25], which identifies HSCO as a superior electrolyte for future device applications. It is interesting to note that the thickness independent ionic conductivity (**Supplementary Fig. 6**) can nicely ruling out the contribution of surface and film/substrate interface, and attribute the measured results to the intrinsic bulk effect of HSCO through the transportation of proton within the lattice (**Supplementary Figs. 7** and **8** and **Supplementary information section 1**).

Furthermore, the electronic conductivity $\sigma_e$ of HSCO was measured and estimated to be two orders of magnitude smaller than the proton conductivity $\sigma_i$ (**Fig. 1d** and **Method Section**), which is attributed to the enhanced electronic band gap in HSCO due to the suppressed Co-O hybridization (**Supplementary Figs. 10** and **11** and **Supplementary** **information section 2**). We note that such distinct difference between ionic and electronic conductivity is essential to prevent the occurrence of electronic current leakage along with ionic transportation.

As shown in **Fig. 1b**, the high proton conductivity and low activation energy in the HSCO can be attributed to the intrinsically enhanced proton concentration as well as structurally ordered oxygen vacancy channels. To reveal the proton diffusion process in HSCO, we studied the Pt-catalyzed protonation process of SCO in forming gas (**Fig. 2a** and **b**). During the study, the SCO thin films were partially covered with noble metal (Pt) stripes with the interval spacing of 300 μm (**Fig. 2c**). It is interesting to observe that despite of the large lateral spacing, which corresponds to the diffusion length of proton, the sample changes rapidly from SCO into HSCO after short period (~240 seconds) of annealing process at 100 ºC, as indicated by the change of transparency (**Fig. 2c**) through bandgap modulation (**Supplementary Fig. 1**) [18] as well as distinct XRD diffraction peaks (**Fig. 2d**). We note that catalysis induced HSCO samples show very similar properties to those processed by the ILG (**Supplementary Fig. 1** and **2**).

To obtain further insights for the proton diffusion, we carried out *in-situ* XRD studies (**Supplementary Fig. 12**) during the pronation induced phase transformation. To the first-order approximation we can take the phase transformation time as the diffusion duration. This estimation assumes that the phase transformation is dominated mainly by the horizontal diffusion process since the sample thickness is about three orders of magnitude smaller than the lateral diffusion length. We note that this measurement would provide a lower-bound estimation of the diffusion coefficient ($D$) following the relation of $L_D = \sqrt{Dt}$, where $L_D$ and $t$ represent the diffusion distance (half of the interval spacing for stripes, which is 150 μm in this study) and diffusion time, respectively. Thus, the lateral diffusion coefficient at 100 ºC can be estimated as $9.09 \times 10^{-7}$ cm$^2$/s, which is much higher than that for HSNO ($1.6 \times 10^{-7}$ cm$^2$/s) at 300 ºC (Ref. 14). The diffusion coefficients at other temperatures were also obtained in the same manner and then summarized in **Fig. 2e**. For comparison, we also calculated the proton diffusion coefficient from the measured proton conductivity (shown in **Fig. 1d**) according to the relationship [14] of $\sigma = \frac{c(Ze)^2 D}{k_B T}$, where $c$ is the concentration of protons within the lattice, $Z$ is the number of electron charge for each ion carrier, and $e$ is the electron charge. Remarkably, the diffusion coefficients obtained by these two methods agree quite well (**Fig. 2e**), which further suggests that the experimentally observed proton conductivity is indeed the intrinsic nature of the HSCO.

We note that the SCO thin film epitaxially-grown on LSAT substrate with compressive strain shows the horizontally ordered oxygen vacancy channels; while the SCO grown on $DyScO_3$ $(110)_o$ with tensile strain shows the vertically ordered oxygen vacancy channels [26]. Thus the comparison between these two cases would provide us the direct information about the correlation between the proton diffusion and the orientation of vacancy channels. Despite the rapid phase transformation observed in the former one, the transformation driven by in-plane proton diffusion in the latter case however cannot be completed even with extended time (**Supplementary Fig. 13**), which strongly suggests that the proton diffusion process is indeed anisotropic and dominated by the orientation of the ordered oxygen vacancy channels.

To obtain further insights of the prominent proton conductivity in the HSCO, we performed the first-principles calculations to investigate the corresponding hydrogen transport pathway (**Method Section** and **Fig. 3**). **Figure 3a** and **3b** shows sequential snapshots of the calculated crystalline structures with an introduced extra hydrogen atom (denoted as $H^9$) (blue sphere) transporting along the oxygen vacancy channel viewed along $c$ axis and $a$ axis, respectively. Throughout the whole process, $H^9$ is strongly confined within the tetrahedral layers with smooth evolution paths. When projected into the $a$-$b$ plane (**Fig. 3c**), it can be clearly seen that $H^9$ transports mainly with the assistance of equatorial oxygen atoms at such layer. During the transportation, $H^9$ first bonds with the equatorial $O^1$ (step ①), and then through the elastic deformation of the oxygen tetrahedron, the $O^1$-$H^9$ bond diffuses forward together (step ②). When approaching an adjacent equatorial $O^2$ at the same atomic layer, $H^9$ is exchanged from $O^1$ and bonded to $O^2$ through transient hydrogen bonding (steps ③ and ④). With the same manner (steps ⑤ to ⑧), $H^9$ travels forward to bond with the next equivalent $O^1$ and finishes one transportation cycle (i.e. travelling one unit-cell). A dynamic process during the proton transportation can be better viewed in the **Supplementary Videos 1-3** along the three different crystalline directions. **Figure 3d** displays the calculated energy profile along the proton transportation path with two series of energy barriers, in which the broad peaks around 0.12 eV are corresponding to the deformation of the oxygen tetrahedron, while the pronounced peaks around 0.5 eV stem from the hydrogen exchange between equatorial oxygen atoms $O^1$ and $O^2$. We note that only single-ion transportation was considered in this calculation, while the potential barrier might be further reduced in the case of multi-ion concerted migration[27], as demonstrated theoretically for the Li ion diffusion. Nevertheless, the current calculation results provide qualitatively a rational atomic-scale picture for the observed enhanced proton conductivity.

Finally, to prove the practical feasibility of the HSCO electrolyte, a proof-of-concept lateral dual-chamber SOFC was designed with the schematic diagram shown in the inset of **Fig. 4a**, in which the anode (Pd) and cathode (Pt) are gaseously separated into different chambers with a thin layer of PTFE, while they are still electrically (ionically) connected through the HSCO electrolyte. As shown in **Fig. 4a**, when $H_2$ and $O_2$ gas were simultaneously introduced into the corresponding chambers, this device shows a rapid and steady response with output of notable open circuit voltage (OCV) of ~0.57 V, indicating a successful conversion of the chemical energy into electricity. **Figure 4b** shows the temperature dependent SOFC performance, from which the fuel cell function can be verified from room temperature to 120 ºC. The weak correlation between OCV and temperature further suggests that the OCV is indeed due to the intrinsic process. As control experiment, a cell using as-grown SCO as electrolyte was also tested, which however fails to deliver any output at the same temperature region. Thus, the successful demonstration of the fuel-cell working principle identifies the newly discovered HSCO as an excellent proton electrolyte with broad range of applications. Furthermore, we envision that the reported protonated brownmillerite structure opens up a new avenue to engineer novel solid oxide electrolyte through rational design.

**Figures and Captions**

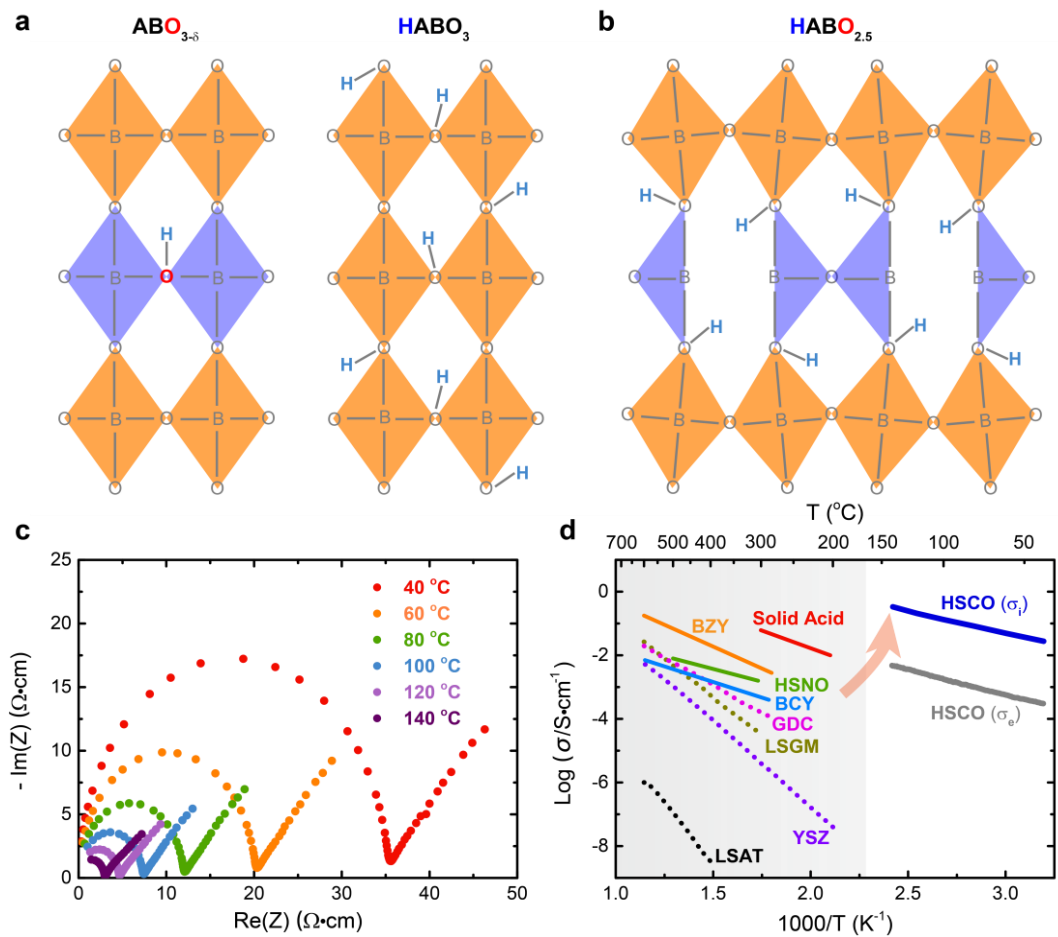

**Fig. 1 | Design philosophy for colossal proton conductivity at HSCO electrolyte. a,** Schematic illustrations of solid oxide proton electrolytes with oxygen vacancy mediated protonation (ABO$_{3-\delta}$) and direct-protonated perovskite oxide (HABO$_3$), respectively. **b,** Proposed crystalline structure for enhanced proton conductivity based on the protonated brownmillerite structure HABO$_{2.5}$. **c,** Temperature dependent Nyquist plots of the impendence spectra for HSCO measured at the atmosphere of forming gas (H$_2$ : Ar =10: 90). **d,** Comparison of proton conductivity in HSCO electrolyte with other solid oxide electrolytes, in which $\sigma_i$ (blue) and $\sigma_e$ (gray) represent ionic and electronic conductivity, respectively. A series of oxygen (LSGM- La$_{0.8}$Sr$_{0.2}$Ga$_{0.8}$Mg$_{0.2}$O$_3$, GDC-Ce$_{0.8}$Gd$_{0.2}$O$_{1.9-\delta}$, YSZ-(ZrO$_2$)$_{0.9}$(Y$_2$O$_3$)$_{0.1}$) (Refs. 21 and 22) and hydrogen (BZY-BaZr$_{0.8}$Y$_{0.2}$O$_{3-\delta}$, HSNO-HSmNiO$_3$, BCY-BaCe$_{0.8-x}$Zr$_x$Y$_{0.2}$O$_{3-\delta}$) (Refs. 11, 14 and 23) ionic electrolytes were itemized as reference. LSAT refers to the oxygen ionic conductivity of the substrate LSAT (001) used for growing SCO thin films.

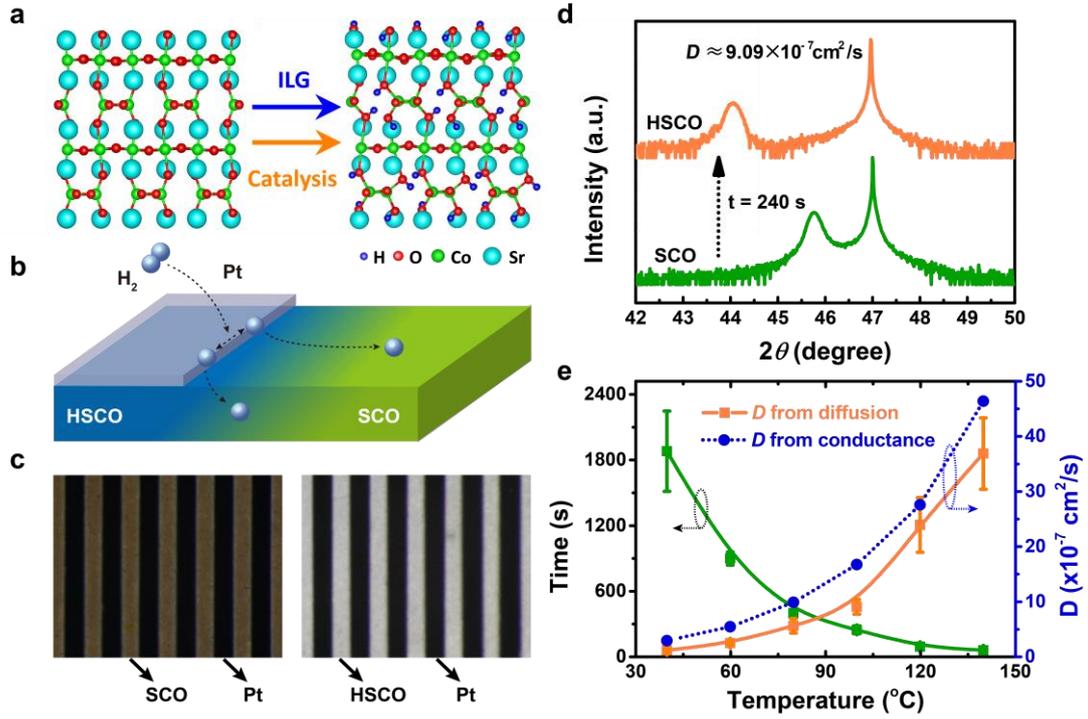

**Fig. 2 | Noble metal catalysis induced protonation in strontium cobaltite thin films. a,** Structural demonstration of the phase transformations realized by both ILG and noble-metal catalysis. **b,** Schematic illustration of the lateral proton diffusion induced protonation in SCO thin films. **c,** Comparison of optical images of as-grown SCO thin film (left panel) on both side polished LSAT (001) substrate with Pt stripes (with the interval spacing of ~300 μm) and that after annealed at 100 °C in forming gas (right panel). **d,** Comparison of the X-ray diffraction $\theta$-$2\theta$ scans between the as-grown SCO and the noble-metal catalysis (at 100 °C) induced HSCO. **e,** Summarized temperature dependent phase transformation times as well as the corresponding estimated proton diffusion coefficients at a series of temperatures. For the comparison purpose, the diffusion coefficient $D$ was also calculated from the proton conductivity shown in **Fig. 1d,** according to the equation of $\sigma = \frac{c(Ze)^2 D}{k_B T}$.

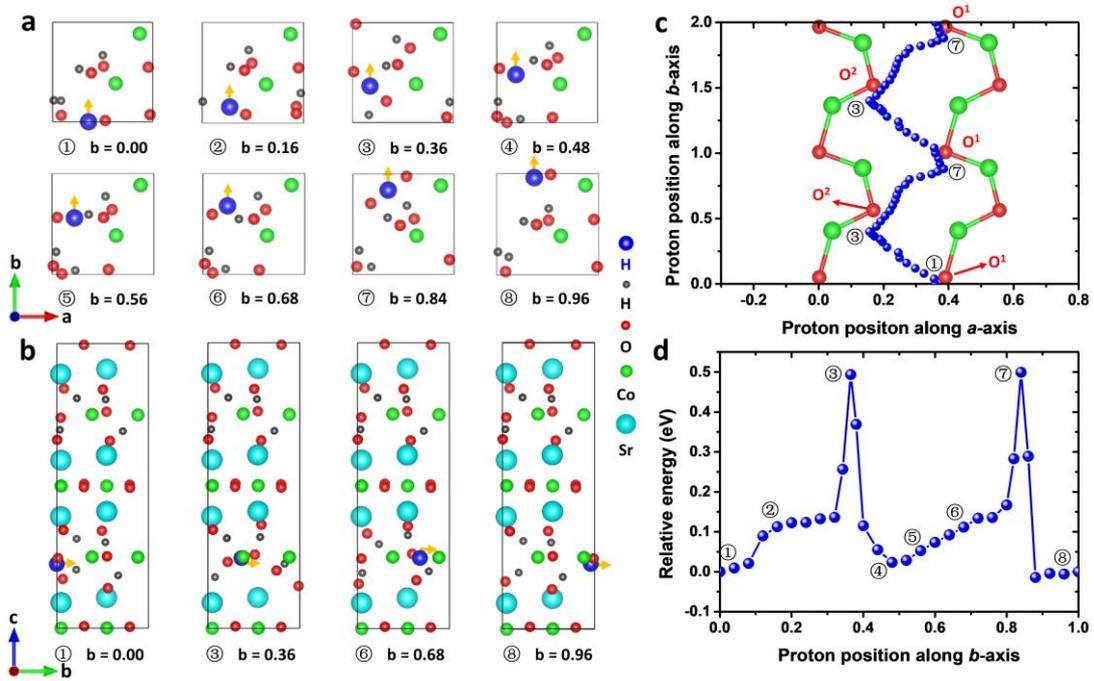

**Fig. 3 | Theoretical calculated hydrogen diffusion path within the HSCO. a-b,** Sequential snapshots of calculated HSCO crystalline structures during the hydrogen transportation along the ordered oxygen vacancy channels (*b* axis). The crystalline structures are viewed along [001] (or *c* axis) and [100] (or *a* axis) directions respectively, and the transporting hydrogen is highlighted as blue sphere. The parameters b used in the calculation corresponds to the hydrogen displacements (ratio to the lattice constant) along the [010] direction (or *b* axis). **c,** Projection of hydrogen transporting trajectory within (001) plane, in which the proton travels mainly along the oxygen vacancy channels. $O^1$ and $O^2$ refer to two neighbored equatorial oxygen atoms within the tetrahedral layers. **d,** Calculated energy potential profile for hydrogen transportation within the vacancy channels of the HSCO. Symbols ① to ⑧ in (**c**) and (**d**) correspond to the sequential snapshots shown in (**a-b**).

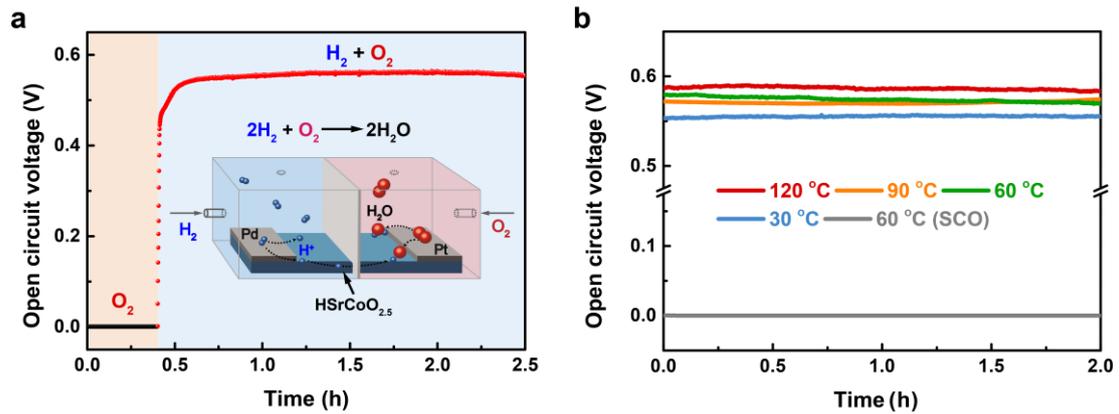

**Fig. 4 | Proof of concept planar dual-chamber solid oxide fuel cell with the HSCO as proton electrolyte. a,** Environmental atmosphere dependence of the open circuit voltage (OCV) for a typical dual-chamber SOFC. While the device in pure $O_2$ atmosphere shows no signature of OCV, the OCV turns on rapidly once the hydrogen fuel is further introduced, and then remains persistence during the operation. Inset shows the schematic diagram of the newly designed dual-chamber fuel cell setup, in which the lateral HSCO thin film acts as the proton electrolyte, while the Pd and Pt metals in different chambers are employed as the anode and cathode, respectively. **b**, Temperature dependent open-circuit voltage for a lateral dual-chamber fuel cell.

**Acknowledgements** This study was financially supported by the Basic Sciencee Center Project of NFSC under grant No. 51788104; the National Basic Research Program of China (grants 2015CB921700 and 2016YFA0301004); the National Natural Science Foundation of China (grant 51561145005); the Initiative Research Projects of Tsinghua University (grant 20141081116); and the Beijing Advanced Innovation Center for Future Chip (ICFC).

**Author contributions** P.Y. conceived the project and designed the experiments. N.L., Y.J.W., Z.L. and J.Z. fabricated the thin films. N.L., Y.J.W. and H.L. carried out ILG, structural, optical and compositional characterization. N.L. performed the impedance spectroscopy, direct current measurement, and double chamber fuel cell characterization and test, with H.L., D.Z., Y.J.W., S.Y., M.W. and J. M. S. Q. performed the theoretical analysis and calculations, under the supervision of J.W. Q. H., Y. J. W., J.G. and K.D. performed the soft X-ray absorption measurements. J. Z. Z, S. T., J. Y. Z. Y. T. and C.N. discussed the results. N.L. and P.Y. wrote the manuscript, and all authors discussed results and commented on the manuscript.

**Author Information** The authors declare no competing financial interests. Correspondence and requests for materials should be addressed to P.Y. (yupu@tsinghua.edu.cn) and J.W. (wu@tsinghua.edu.cn).

**Methods**

**Films growth and characterization.** Thin films were grown by a customized reflection high-energy electron diffraction (RHEED) assisted pulsed laser deposition system. Epitaxial $SrCoO_{2.5}$ (SCO) thin films were grown on $(LaAlO_3)_{0.3}(SrAl_{0.5}Ta_{0.5}O_3)_{0.7}$ (LSAT) (001) and $DyScO_3$ (110)$_o$ substrates at 750 °C in 100 mTorr of oxygen pressure. The laser energy (KrF, $\lambda$ = 248 nm) was fixed at 1.2 J/cm$^2$ with a repetition rate of 2 Hz. After growth, the samples were cooled down to room temperature with 100 mTorr oxygen pressure at a cooling rate of 7 °C/min. The crystalline structures of thin films were characterized by a high-resolution four-circle X-ray diffractometer (Smartlab, Rigaku).

**Protonation of the strontium cobaltite.** Protonation of the strontium cobaltite thin films can be realized through two methods, i.e., ionic liquid gating (ILG) and noble metal (Pt or Pd) catalysis [14, 18]. For the former case, the hydrogen ions are obtained through the electrolysis of trace water within the ionic liquid, as discussed in our previous studies [18]. **Supplementary Figure 1a** presents the *in-situ* XRD studies during the ILG with gating voltage of +3.3 V, in which the proton is inserted into the as-grown SCO to form corresponding protonated $HSrCoO_{2.5}$ (HSCO). During the protonation, the pseudo-cubic c-axis lattice constant is significantly expanded from 3.97 Å to 4.11 Å, while the in-plane lattice is constrained by the LSAT (001) substrate. These results suggest that the high crystalline nature of thin films is retained, and without any detectable deterioration. For the latter case, the protonated SCO thin films were achieved through the forming gas ($H_2$: Ar = 1: 9) annealing under the catalysis of noble metal. For this purpose, the SCO thin films were covered by patterned thin layer (~10 nm) of Pt (or Pd) nanoparticles deposited through sputtering. Once the forming gas is introduced into the sample atmosphere environment, the hydrogen molecular can be catalyzed into active free radicals through the catalysis process, which would intercalate into the crystalline lattice of SCO to form the corresponding protonated HSCO phase through diffusion.

**Identification of the protonated HSCO phase.** In order to verify the protonation process as well as the corresponding product, a series of detailed measurements including second ion mass spectroscopy (SIMS), energy dispersion X-ray spectroscopy (EDS), UV-VIS-NIR and soft X-ray absorption (sXAS) were carried out. For the noble metal catalyzed HSCO, the noble metal layer was removed by plasma etching before the SIMS, optical spectroscopy and sXAS measurements. Through the SIMS measurements, the average hydrogen concentration within the ILG protonated $H_xSrCoO_{2.5}$ is estimated as ~$1.72\times10^{22}$ atom/cm$^3$ (x = 1.06 ± 0.06), while for the case of noble metal catalysis, the hydrogen distribution is less uniform with a lightly higher proton concentration of ~$2.27\times10^{22}$ atom/cm$^3$ (x = 1.4 ± 0.26) (**Supplementary Fig. 1b**). We note that both values are well above the limitation of the oxygen vacancy concentration in conventional oxides. Meanwhile, the negligible loss of oxygen element was also confirmed through the EDS analysis (**Supplementary Fig. 1c**). The band gaps revealed by optical absorption measurements reach up to ~3.0 eV (ILG) and 3.6 eV (Catalysis) in the HSCO phase as compared with ~2.0 eV in the as-grown SCO phase (**Supplementary Fig. 1d**). To direct access the valence states of Co ion as well its hybridization with the oxygen ion, we employed sXAS to study the Co *L*-edge and O *K*-edge of SCO and HSCO samples (**Supplementary Fig. 2**). Through comparison, a notable shift of the Co $L_3$ edge to lower energy can be clearly identified in both HSCO, which suggests that the $Co^{3+}$ ions in pristine SCO have switched into $Co^{2+}$ in the HSCO phase [18] through the protonation induced electron doping. Furthermore, the O-Co *p-d* hybridization in HSCO is also largely suppressed, as indicated by the disappearance of pre-edge at the O *K*-edge absorption spectra, which suggests that the Co ions become highly ionic and lead to the enlarged bandgap. We note that the enhanced electronic bandgap indicates a suppression of electronic conductivity in HSCO, which is essential for a superior ionic conductor.

**Measurement of the proton conductivity in HSCO.** The proton conductivity was measured by a current (AC) impendence meter (Agilent E4980A) with the frequency range of 20 Hz to 2 MHz. The amplitude of AC voltage employed was set at 200 mV throughout the measurements, while identical results can be obtained with voltage of 50 mV as well, from which the linear response nature of the electrical component can be verified. To prepare the device structure for the ionic conductivity measurements, the HSCO were first obtained by ILG. Then the Pt (or Pd) electrode was sputtered on the protonated HSCO thin films to form corresponding patterns with hard mask. Subsequently, the AC complex impendence spectroscopy in the form of Nyquist plot was collected in forming gas (with $H_2$: Ar volume ratio of 10: 90 and total flow rate of 100 sccm) or other atmosphere gases at variable temperatures. The fabrication and measurement process of the device are demonstrated in the left panel of **Supplementary Fig. 3** as well as the inset of **Supplementary Fig. 5b**. The measurements were carried out at the temperature up to 140 $^o$C, due to the fact that the HSCO would transform back into SCO phase with temperature higher than 160 $^o$C as revealed by **Supplementary Fig. 4**.

**Fitting of the complex impedance spectroscopy**. The complex impedance spectra can be fitted with a model circuit consisting the following elements: parallel sample resistance (R2) and constant phase element (CPE1), in series with an intrinsic resistance (R1) in the circuit and a Warburg-type impedance representing the electrode polarization feature (**Supplementary Fig. 5a**). For illustration purpose, a typical impedance spectrum obtained at 60 $^o$C were presented together with the corresponding fitting in **Supplementary Fig. 5b**. The complex impedance data shows a straight-line at low frequency region and a semicircle at high frequency. The slope of the straight-line approximately equals to 1, which represents the polarization of the electrodes. The sample resistivity or the proton conductivity can be obtained from the intersection of the semicircles with the real axis.

**Measurement of the electronic conductivity in SCO and HSCO.** We note that in conventional ionic conductors, their ionic conductivity as well as type of charge carriers would be very sensitive to the measurement gaseous environments, and the change of carrier would lead to strong modified activation energy [11]. Our measurements on SCO at different gaseous environments (Ar, $O_2$ and $H_2$/Ar) reveal almost identical temperature dependent conductivities and activation energy (0.24 eV) (**Supplementary**

**Fig. 9**), at the measurement temperature region from room temperature to 140 ºC. With this, we further probed the protonic and oxygen ionic conductivity of SCO by using it as the electrolyte during the dual-chamber fuel cell operation, which however cannot deliver any voltage output in the same temperature region (**Fig. 4b**), clearly suggesting the protonic and oxygen ionic conductivity are negligible at the measured temperature region. Furthermore, the previously theoretical study of the oxygen ionic transport in SCO revealed a much larger activation energy (larger than 0.62 eV) [28], and experimentally the SCO can only show oxygen reduction reaction at the temperature above 500 ºC (Ref. 26). Therefore, we can assure that the measured conductivity in SCO at the low temperature region (up to 140 ºC) should be attributed its conducting electrons. We note that the HSCO shows almost identical conductivity at different gaseous environments of Ar, $O_2$ and $H_2$/Ar, which make it inappropriate to probe its electronic conductivity with the selection of gas environments as general employed in the conventional proton electrolyte. Due to the facts that the electronic conductivity of SCO layer (~0.1 S·cm$^{-1}$ at room temperature) is much larger than that of the HSCO, while it is totally insulating for the proton transportation, we can probe the electronic conductivity of HSCO by measuring a sample with HSCO and SCO in series as shown in **Supplementary Fig. 3h**. Accordingly, the DC measurement across this device would provide a good estimation of the intrinsic electronic conductivity in HSCO, which is estimated to be $3\times10^{-4}$ S·cm$^{-1}$ at room temperature, being more than two orders of magnitude smaller than that of the SCO.

**Mechanism of the enhanced proton conductivity.** To theoretically study the crystalline and electronic structures of HSCO as wells its corresponding hydrogen diffusion process, we employed first-principles calculations using Vienna Ab-initio Simulation Package (VASP) within density functional theory (DFT). The generalized gradient approximation of Perdew-Burke-Ernzerh (GGA-PBE) and the projector augmented wave (PAW) method were carried out to describe the electron-electron interaction and electron-ion interaction, respectively. Plane-wave cutoff energy as 500 eV, dense Monkhorst-Pack k meshes as 7×7×3, and energy self-consistency precision as $10^{-5}$ eV were selected to obtain well converged results. We involved a G-type antiferromagnetic state and a +U correction on cobalt d electrons during the calculations, with on-site Coulomb interaction of U = 4.5 eV and on-site exchange interactions of J = 0.5 eV.

To investigate the hydrogen diffusion process, we constructed a brownmillerite $HSrCoO_{2.5}$ unit cell containing 44 atoms (with 8 H atoms, 8 Sr atoms, 8 Co atoms and 20 O atoms) and then introduced the 9$^{th}$ H ($H^9$) atom and monitored its diffusion path along the *b*-axis within the tetrahedral sublayers. We first constructed a series of crystalline structures with the $H^9$ atom distributing randomly within the system. Due to the presence of substrate confinement, structure relaxations were performed without lattice transformation. Accordingly, we obtained two energetically favorite structures with $H^9$ bonded with two inequivalent equatorial oxygen atoms (i.e. $O^1$ and $O^2$ as shown in **Fig. 3c**) at the tetrahedral sublayers. To mimic the hydrogen transportation path between these two structures, constrained optimizations were performed with a series of sequential displacement of the $H^9$ from the $O^1$ site. With this, we can then capture the completed picture of structural evolution with the corresponding energy barrier profile during the hydrogen transportation.

**Performance test of the designed lateral double chamber full solid oxide fuel cells.** The open-circuit voltages (OCV) of the lateral dual-chamber fuel cell were measured by Keithley 2000 and Zennium-Pro electrochemical workstation (Zahner scientific instrument). The environment temperature was controlled by the reactive chamber of Linkam HFS600-PB4 station. The gases introduced into the separated chamber are forming gas ($H_2$: Ar = 10: 90) as fuel and oxygen as oxidant with flow rate of 90 and 10 sccm, respectively. The inset of **Fig. 4a** displays a schematic diagram of the planar dual-chamber fuel cell, in which the protonated HSCO and noble metals (Pd and Pt) behave as proton electrolyte and electrodes, respectively.

# Supplementary Materials for

# A protonated brownmillerite electrolyte for superior low-temperature proton conductivity

Nianpeng Lu[1†], Yujia Wang[1†], Shuang Qiao[1†], Hao-Bo Li[1†], Qing He[2], Zhuolu Li[1], Meng Wang[1], Jingzhao Zhang[3], Sze Chun Tsang[3], Jingwen Guo[1], Shuzhen Yang[1], Jianbing Zhang[1], Ke Deng[1], Ding Zhang[1, 4], Jing Ma[5], Yang Wu[1, 6], Junyi Zhu[3], Yoshinori Tokura[7], Ce-Wen Nan[5], Jian Wu[1, 4*] and Pu Yu[1, 4, 7*]

[1]State Key Laboratory of Low Dimensional Quantum Physics and Department of Physics, Tsinghua University, Beijing, 100084, China

[2]Department of Physics, Durham University, Durham DH1 3LE, United Kingdom

[3]Department of Physics, The Chinese University of Hong Kong, Shatin, New Territories, Hong Kong, China

[4]Collaborative Innovation Center of Quantum Matter, Beijing 100084, China

[5]State Key Lab of New Ceramics and Fine Processing, School of Materials Science and Engineering, Tsinghua University, Beijing 100084, China

[6]Tsinghua-Foxconn Nanotechnology Research Center, Tsinghua University, Beijing 100084, China

[7]RIKEN Center for Emergent Matter Science (CEMS), Wako 351-198, Japan

†These authors contributed equally to this work.

Correspondence to: P.Y. (yupu@tsinghua.edu.cn) and J.W. (wu@tsinghua.edu.cn).

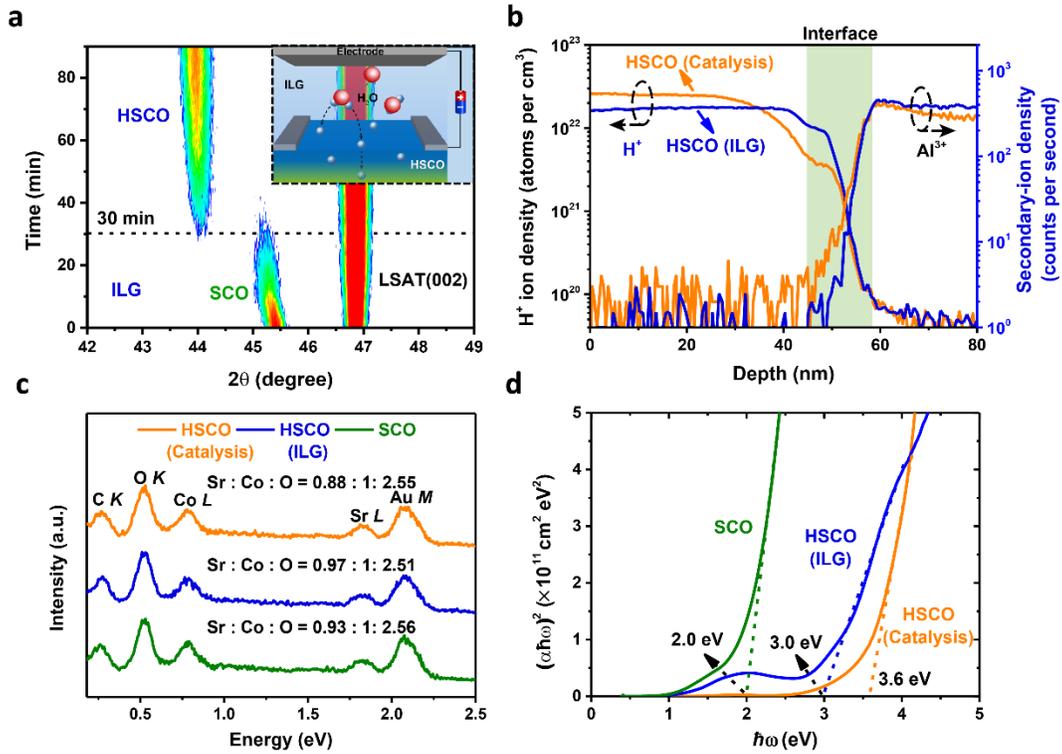

**Supplementary Figure 1 | Identification of HSCO phase. a,** *In-situ* XRD studies of the ILG induced protonation from SCO thin film, from which the phase transformation can be clearly identified from the lattice expansion. Inset shows the schematic illustration of the ILG process. The gating voltage of +3.3 V was employed for this measurement. **b,** Depth profiles of the $H^+$ and $Al^{3+}$ ions from the protonated HSCO samples prepared by ILG and Pt catalysis in forming gas, measured with secondary-ion mass spectrometry. The $Al^{3+}$ signal from the LSAT substrate was employed as a marker for the interface position. **c,** Energy-dispersive X-ray spectroscopy results for as-grown SCO (green) as well as ILG-induced (blue) and Pt catalyzed (orange) HSCO thin films. **d,** Comparison of the optical absorption spectra of the SCO and protonated HSCO phases, from which the direct band gaps can be obtained.

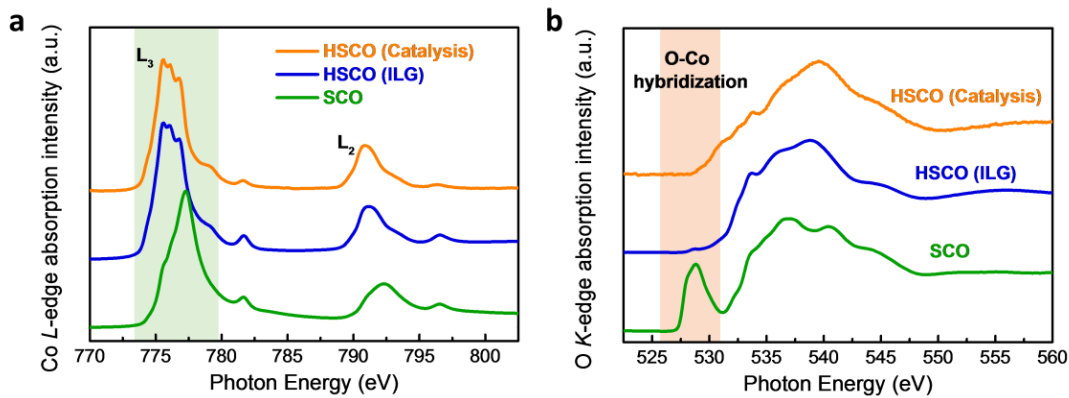

**Supplementary Figure 2 | Determination of the valence states as well as the O-Co hybridization in SCO and HSCO phases.** sXAS of the (**a**) Co *L*-edge and (**b**) O *K*-edge of the pristine SCO and protonated HSCO phases formed through ionic liquid gating (ILG) and noble metal (Pt) catalysis process, respectively.

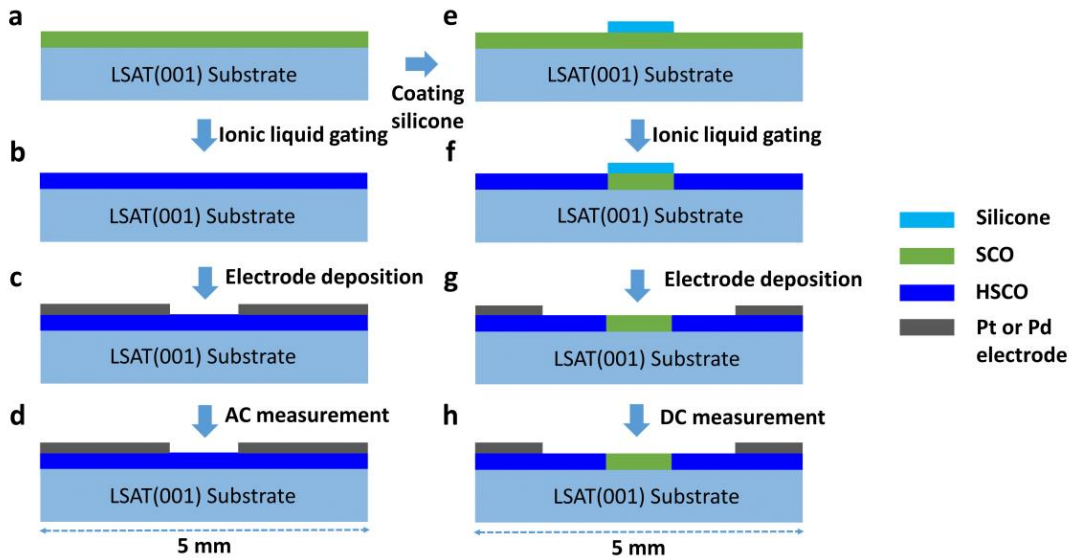

**Supplementary Figure 3 | Schematic device configurations for the ionic (a-d) and electronic conductivity (e-h) measurement along with device fabrication flowchart.** During the electronic conductivity measurement, the pristine SCO phase region was used as block layer for the proton transportation, since the electronic conductivity of SCO layer (~0.1 S·cm$^{-1}$ at room temperature) is much larger than that of the HSCO. Accordingly, the DC measurement across this device would provide a good estimation of the intrinsic electronic conductivity in HSCO, which is estimated to be $3\times10^{-4}$ S·cm$^{-1}$ at room temperature.

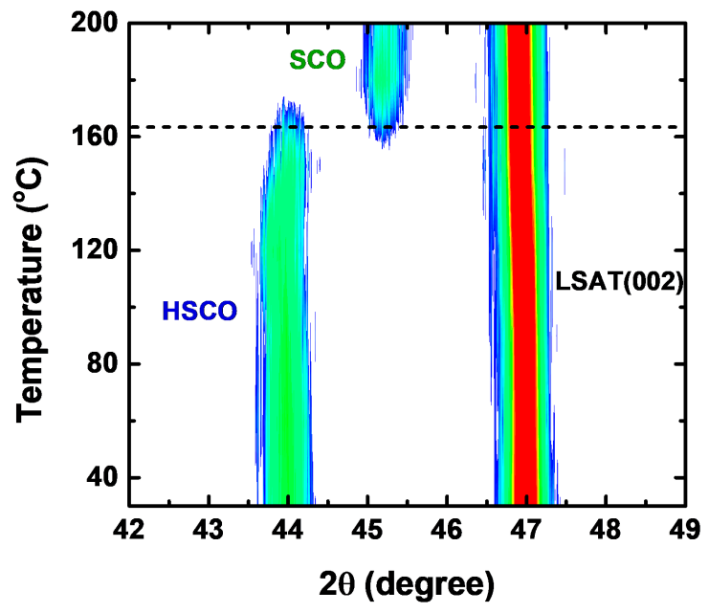

**Supplementary Figure 4 | Thermal stability test for HSCO thin films.** *In-situ* temperature dependent XRD *θ-2θ* scans for HSCO thin film obtained by ILG in the atmosphere of forming gas ($H_2$ : Ar =10: 90). From the measurements, it can be seen that the HSCO phase remains stable with the temperature up to 160 °C, and above which the HSCO transforms back into SCO.

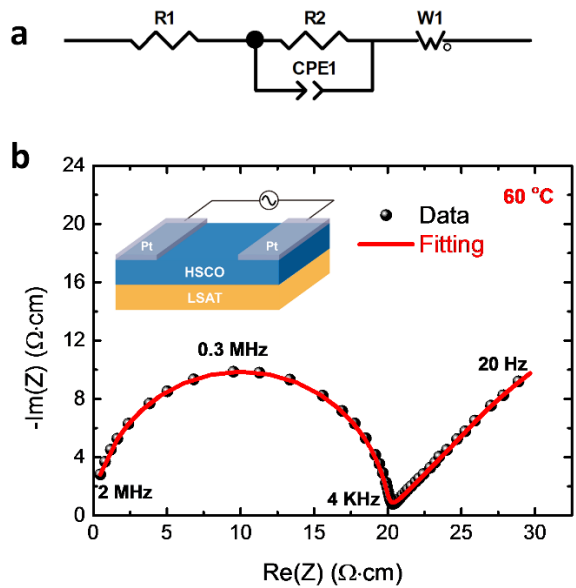

**Supplementary Figure 5 | Equivalent circuit and corresponding fitting for the complex impedance spectra.** The equivalent circuit (**a**) and corresponding fitting (**b**) for the representing Nyquist impedance spectra obtained at 60 °C from the ILG induced HSCO sample. Inset in (**b**) shows schematic diagram of device architecture for the impedance spectroscopy measurement. From the fitting, the proton conductivity at 60 °C can be estimated as 0.051 S·cm$^{-1}$.

# Section 1: Extended experimental evidences for the enhanced proton conductivity in HSCO.

To prove that the measured ionic conductivity is indeed the intrinsic property of the newly discovery HSCO due to its abundant proton content as well as well-ordered oxygen vacancies, we carried out a series of extended studies as discussed below.

**Ruling out the contribution of ionic liquid residual**. Right after the ionic liquid gating, we used acetone and isopropylalcohol to clean the surface of gated films to remove any possible residual ionic liquid, which can exclude the contribution of the ionic liquid residual at the surface to the measured conductance. Note that since the phase transformation is nonvolatile, both the cleaning and device fabrication procedures will not change the lattice structure.

**Thickness dependent studies.** We carried out the thickness dependent ionic conductivity measurements at samples with thicknesses of 18 nm, 40 nm, 60 nm and 100 nm, as shown in **Supplementary Fig. 6**. It can be clearly seen that for all samples, the corresponding ionic conductivity measurements show almost identical temperature dependent behavior with similar proton conductivities and activation energies. As if the surface conductance is the dominated factor for the measured conductance, we would expect the ionic conductivity to be inversely proportional to the film thickness, which however is not observed here. The rather constant ionic conductivities (inset of **Supplementary Fig. 6**) for different thicknesses indeed provide direct evidence that it is an intrinsic bulk effect of this thin films.

**Gaseous environmental dependent measurements.** In order to further verify the intrinsic mechanism for the ionic conductivity, we carried out the control studies of the impedance spectroscopy for HSCO at different atmosphere gases of Ar, $O_2$ and $H_2/Ar$ (**Supplementary Fig. 7**). In conventional proton electrolytes (e.g. $BaZr_{0.8}Y_{0.2}O_{3-\delta}$ (BZY)) formed with diluted oxygen vacancies, the ionic conductivity is very sensitive to the gaseous environments, which was widely employed to determine the type of transporting carriers[1]. However, the obtained impedance spectra for $HSrCoO_{2.5}$ show almost identical features at different gaseous environments. This seemly surprising result can be well understood by taking into the fact that a great number of protons are already present in the electrolyte, and the proton intercalation process is not required for the ionic transport.

**Comparison of the activation energy among different types of transporting ions.** The measured activation energy in our HSCO is 0.27 eV, which is nicely consistent with the widely reported activation energy (~0.2 - 0.3 eV) for high-quality proton conductors[2, 3], while the activation energies for $O^{2-}$ (Refs. 4 and 5) and $OH^-$ (Refs. 6 and 7) are generally much larger (~0.45 - 1.0 eV) or less (~0.11 - 0.15 eV) than this value.

**Isotopic labeling of the transporting ions.** To determine the type of the conducting ions, we carried out the isotopic labeling experiments during fuel cell operation using gas flow of $D_2$/Ar (5%, 90 sccm) and mixed $^{18}O_2$ and $^{16}O_2$ (10 sccm, with $^{18}O_2/^{16}O_2$ ratio of 1:4), respectively (as shown in **Supplementary Fig. 8a**). To exclude the direct ionic exchange from the surface of HSCO electrolyte, the majority of HSCO surface is covered with a thick layer of silicone. For comparison purpose, we prepared two identical fuel cells operated (for two hours) with full open circuit voltage (OCV) and 1/2 OCV outputs, respectively at 60 °C. We note that for the cell with 1/2 OCV operation, the device shows energy conversion with clear ionic transport, while for the full OCV case, the ionic transport will be strongly suppressed. Therefore, we should be able to obtain the direct information of the transporting ions by carrying out the compositional analysis for the covered HSCO regions. The corresponding SIMS measurements at the covered regions show clearly enhanced deuterium signal for the cell operating at 1/2 OCV as compared with the one operating at full OCV (**Supplementary Fig. 8b**), while on the other hand the $^{18}O$ composition is almost the same between these two cases (**Supplementary Fig. 8c**), which should be attributed to the nature isotropic abundance of $^{18}O$ (0.2%). Therefore, these measurements show strong and direct evidence that the transporting ions are indeed the proton, but not the oxygen ion and $OH^-$ group.

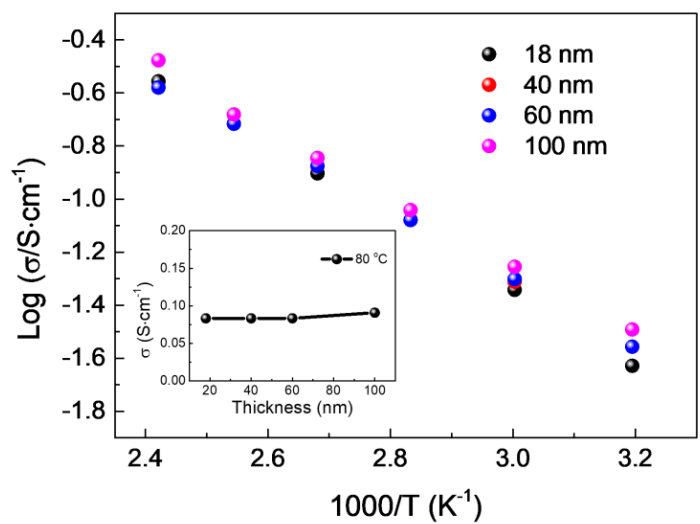

**Supplementary Figure 6 | Thickness dependent measurements of the proton conductivity.** Inset shows the proton conductivity with the different film thicknesses measured at 80 °C. It can be clearly seen that for all samples, the corresponding ionic conductivity measurements show almost identical temperature dependent behavior with similar proton conductivities and activation energies.

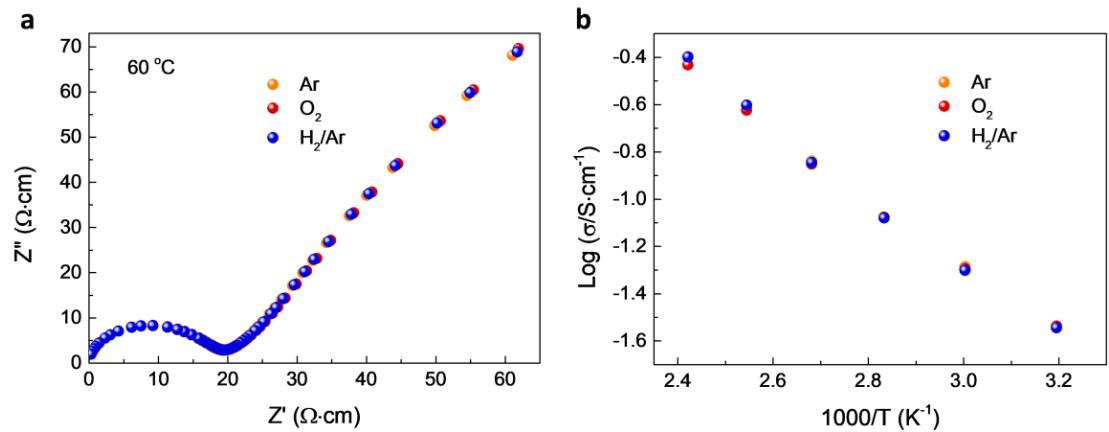

**Supplementary Figure 7 | Atmosphere dependence of the ionic conductivity.** (**a**) Represented impedance spectra and (**b**) temperature dependent proton conductivities for HSCO measured at different gaseous environments of Ar, $O_2$ and $H_2/Ar$, respectively.

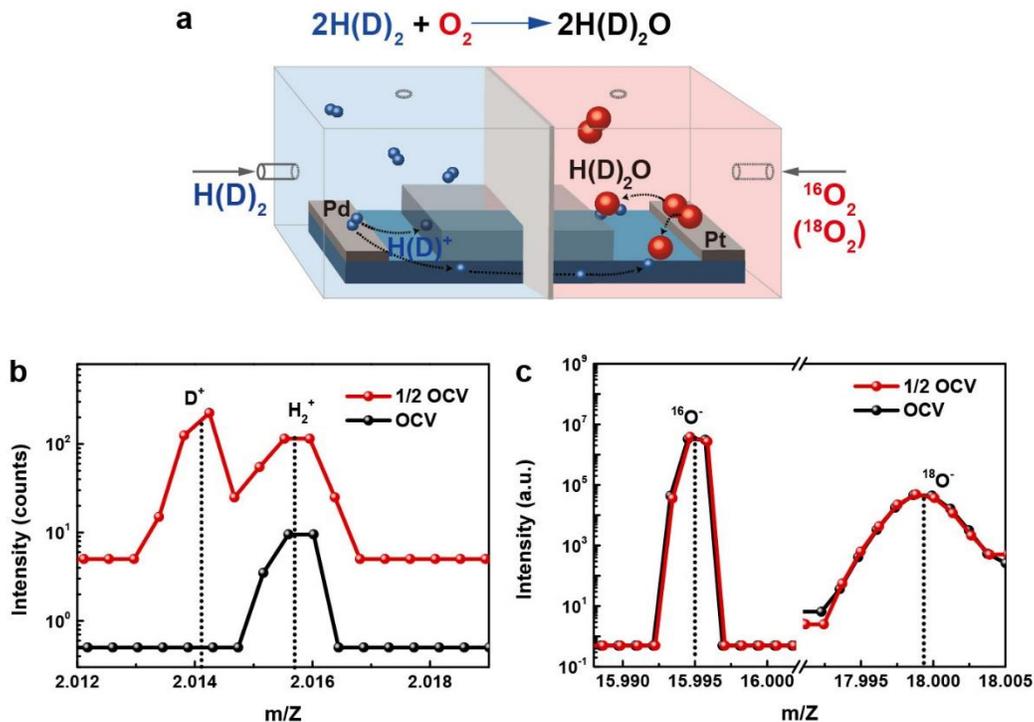

**Supplementary Figure 8 | Isotopic labeling of transporting ions during fuel cell operation.** (**a**) Schematic illustrations of a lateral dual-chamber fuel cell. The HSrCoO$_{2.5}$ electrolyte is covered with thick (~1 mm) silicone (in gray color) to avoid direct ionic exchange through its surface, while the Pd/HSCO and Pt/HSCO interfaces are directly exposed to the forming gas and oxygen gas respectively. Compositional analysis of (**b**) deuterium and (**c**) $^{18}$O at the covered electrolyte regions for fuel cell operation at full open circuit voltage and with load (half open circuit voltage).

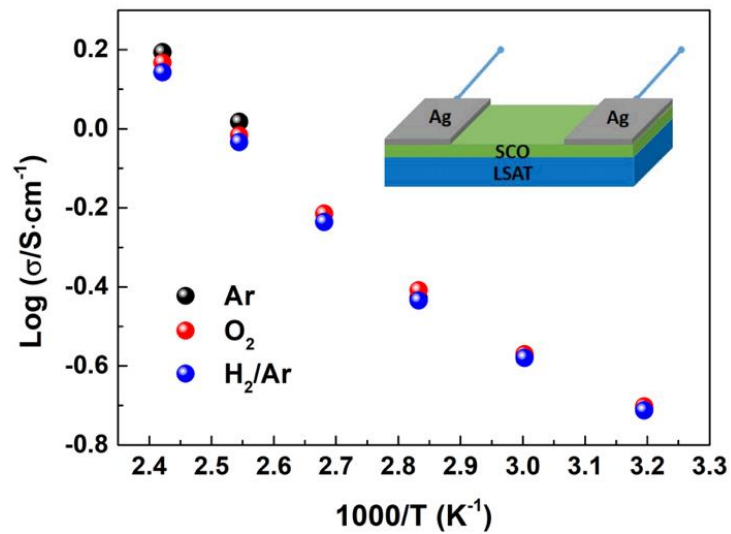

**Supplementary Figure 9 | DC measurement of SCO under different atmosphere condition.** To make sure that the proton will not be purged into the SCO, the DC measurements was performed with Ag electrode, which has negligible hydrogen catalysis capability.

# Section 2: Theoretical discussions of the enhanced electronic band gap in protonated HSCO.

To reveal the mechanism of the enhanced electronic band gap in HSCO, we obtained the projected density of states (PDOS) for Co 3d-orbitals and O 2p-orbitals in both SCO and HSCO system through DFT calculations, as shown as **Supplementary Fig. 10**. When projecting the DOS of Co atoms into their $e_g$ and $t_{2g}$ orbitals, we find that both $e_g$ and $t_{2g}$ orbitals overlap with each other among a large range of energy spectra (nearly from -7.5 eV to 2.5 eV), which suggests a strong orbital hybridization between them, and therefore these orbitals are no longer the good basis set for the wave-functions. With that, it would not be appropriate to draw directly the 3d-orbital diagrams in the current system as well-employed in the study of perovskite oxides. Instead, we would like to use the PDOS to discuss the intrinsic nature of the band gap enhancement through protonation. To accord with the crystalline structures of both SCO and HSCO, we define the Co ions as $Co^O$ (at octahedral layers) and $Co^T$ (at tetrahedral layers). For both SCO and HSCO, the PDOS results show that in $Co^O$ ($Co^T$) the up (down) spin states are almost fully occupied, while a large portion of down (up) spin is empty, implying a high spin state for Co ions in both cases. Moreover, the PDOS on the oxygen p-orbital in SCO shows a partially filled $2p$-orbitals, indicating the electron hybridization between O-$2p$ and Co-$3d$. While for the protonated HSCO, the hybridization is clearly suppressed with reduced unfilled $2p$ state. These calculated results are consistent with our experimental oxygen $K$-edge XAS data (**Supplementary Fig. 2b**).

We note that in the brownmillerite SCO, one might expect the valence states of $Co^{2+}$ and $Co^{4+}$ in tetrahedral and octahedral layers respectively. However this assignment is not consistent with our experimental results, in which both $Co^O$ and $Co^T$ possess almost identical valence state of $Co^{3+}$ in SCO, and could be further reduced into $Co^{2+}$ upon protonation in HSCO (Refs. 8 and 9). Furthermore, theoretical calculation also supports the result by considering the similar total $d$ electrons between $Co^O$ and $Co^T$ in both SCO and HSCO cases (**Supplementary Fig. 10**). Therefore, it is clear that a pronounced charge redistribution would be developed between the tetrahedral and octahedral sublayers likely through the assistant of the interfacial oxygen ($O^I$) located between octahedral and tetrahedral sublayers. Indeed, the partial charge distribution maps of SCO (**Supplementary**

**Fig. 11a**) indicates a stronger intralayer Co-O interaction with the tetrahedral sublayer (i.e. $Co^T$-$O^I$), which is also evidenced by the fact that the $Co^T$-$O^I$ bond distance is 0.44 Å shorter than that for $Co^O$-$O^I$. Such a bond imbalance would naturally suggest that the $O^I$ contributes more to the tetrahedral layer, and accordingly these two Co polyhedral units can be assigned as $[(CoO_{2+\delta})^{2-}]$ and $[(CoO_{3-\delta})^{2-}]$ for the tetrahedra and octahedra, respectively. As the consequence, the Co atoms would hold the same $Co^{3+}$ state (with $\delta \sim 0.5$) for both cases, consistent with our experimental results.

Such a bond modulation scenario can also explain the valence change in HSCO case, in which the inserted H ions locate at the tetrahedral layers and form bonding with the $O^I$ ions. Due to this strong interaction, the weak $Co^O$-$O^I$ covalent bond would be further suppressed as supported by the totally disappeared charge distribution between $Co^O$ and $O^I$ as well as the dramatic shift of $O^I$ ions toward $Co^T$, as confirmed by the partial charge distribution (**Supplementary Fig. 11b**). Specifically, our calculation reveals that the difference between $Co^O$-$O^I$ and $Co^T$-$O^I$ bond length increases from 0.44 Å to 0.81 Å. Therefore, the $O^I$ would couple with the $Co^T$ completely and the actually chemical environment of Co atoms would behave as $[(H_2CoO_3)^{2-}]$ and $[(CoO_2)^{2-}]$ at the tetrahedra and octahedra, and as a result the Co atoms exhibit the same $Co^{2+}$ valence state within the lattice.

With the above discussion, we would be able to build up a scenario with ionic coupling among the charged units of $(H_2CoO_3)^{2-}$, $Sr^{2+}$ and $(CoO_2)^{2-}$ ions in HSCO. Therefore the enhanced band gap in HSCO can be attributed to its enhanced ionic nature. We note that similar phenomenon was also observed in $Li_xFePO_4$ system, where the lithiation process drives the system into a highly ionic nature, and consequently the system opens up a large band gap as well [10].

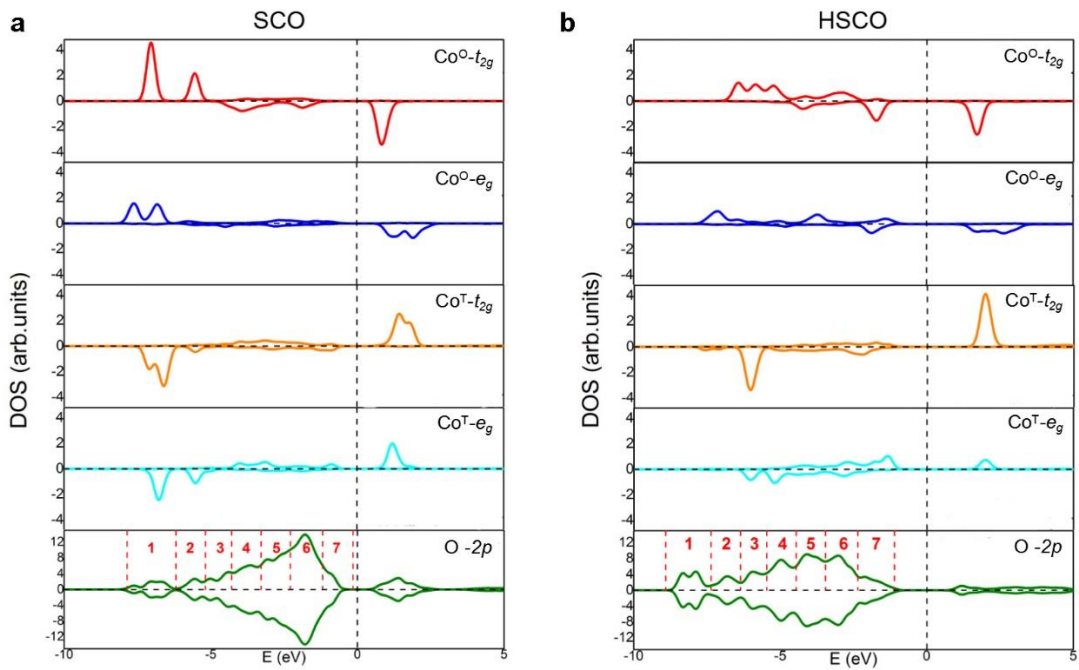

**Supplementary Figure 10 | Projected density of states (PDOS) into Co 3d and O 2p orbitals for (a) SCO and (b) HSCO systems.** The positive and negative values indicate the up and down spin states, respectively, while the magnetic ground states remain to be G-type antiferromagnetism for both SCO and HSCO. The red dash lines in the bottom panel mark different energy windows corresponding to the partial charge distributions presented in **Supplementary Fig. 11**.

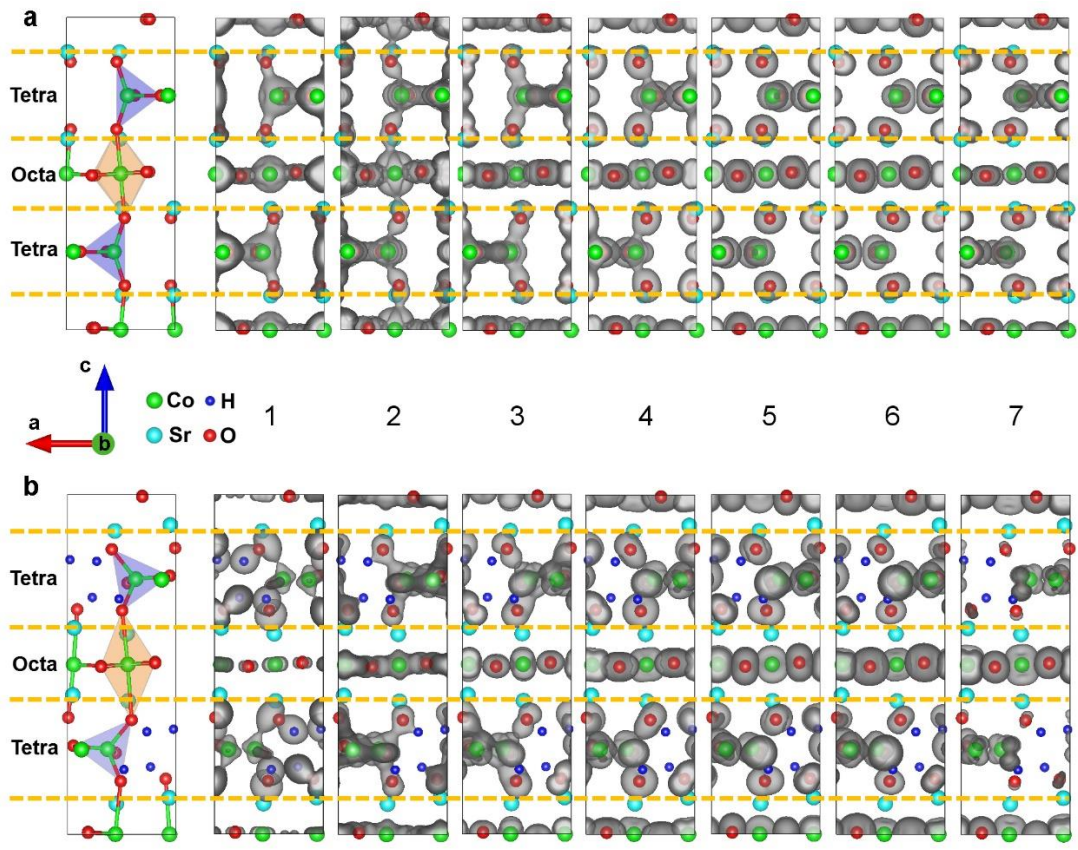

**Supplementary Figure 11 | Partial charge distributions for (a) SCO and (b) HSCO systems.** The different cases (1-7) are corresponding to the energy ranges marked in **Supplementary Fig. 10**. The left panels show the crystalline structures for both SCO and HSCO, in which the transparent blue and orange polyhedrons represent the Co-O tetrahedron and octahedron, respectively. The transparent gray areas in the right panel represent the charge distributions. The yellow dash lines are the guide for different atomic layers within the unit cell.

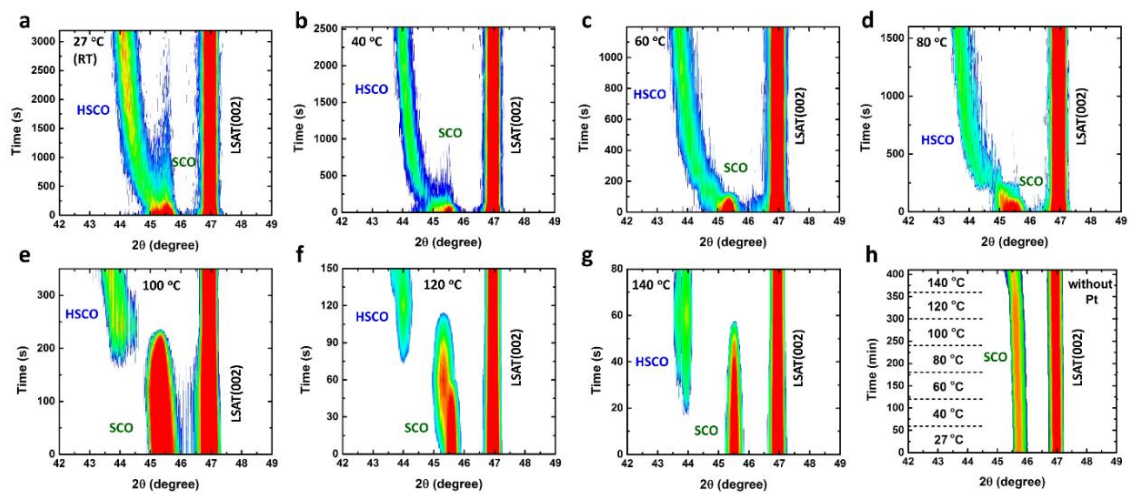

**Supplementary Figure 12 |** *In-situ* **XRD studies during the phase transformation from SCO to HSCO through the Pt catalysis process.** Since the thin films were only partially covered with Pt layer, the phase transformation was thus dominated by the lateral proton diffusion as schematically shown in **Fig. 2b**. The forming temperatures were set at (**a**) 27 ºC , (**b**) 40 ºC, (**c**) 60 ºC, (**d**) 80 ºC, (**e**) 100 ºC, (**f**) 120 ºC and (**g**) 140 ºC, respectively, within the atmosphere of forming gas ($H_2$ :Ar =10: 90). (**h**) For comparison purpose, the SCO thin film without Pt capping was also annealed within forming gas, which shows no detectable phase transformation at the whole temperature region tested.

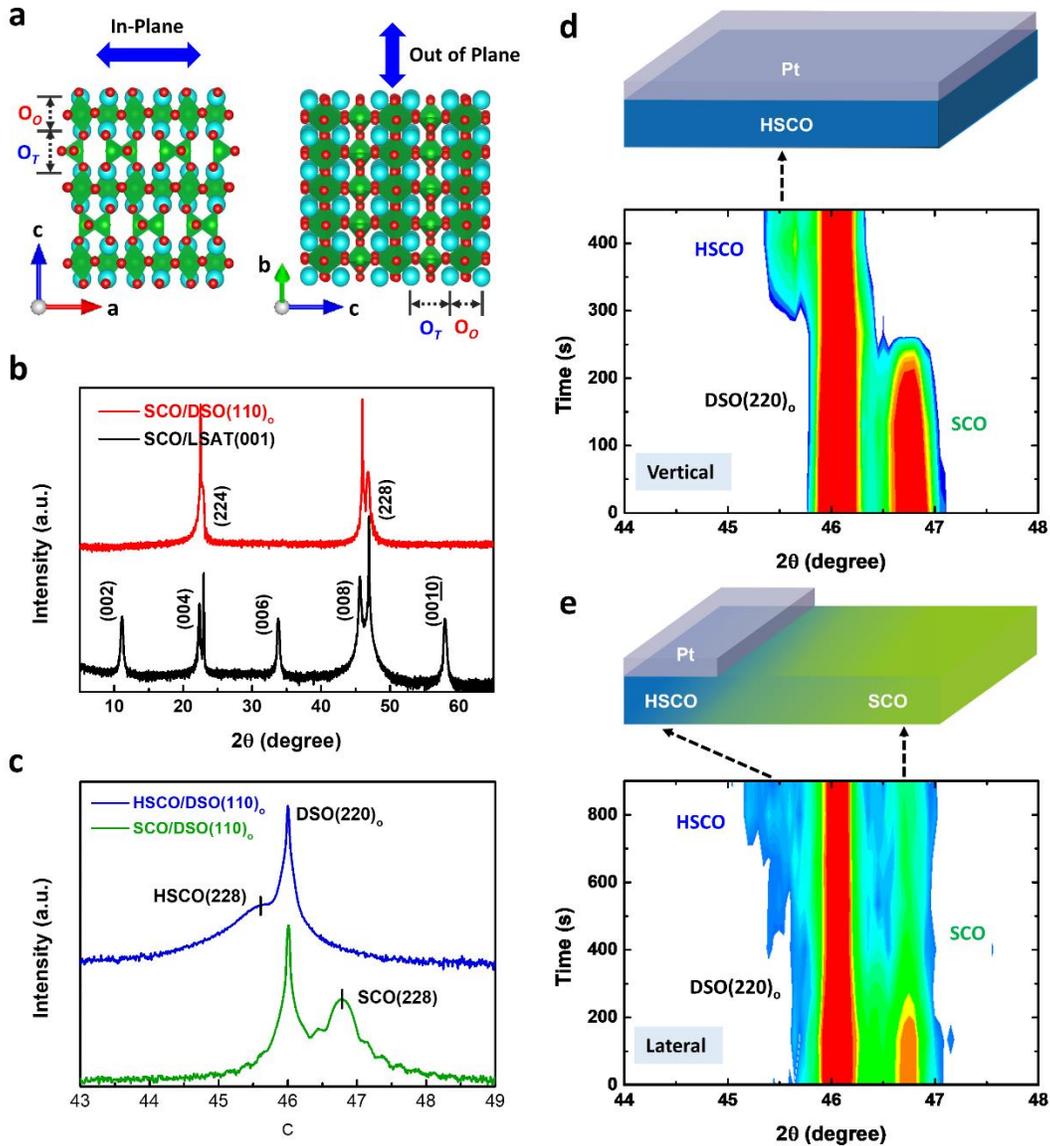

**Supplementary Figure 13 | Correlation between the orientation of oxygen vacancy channel and the hydrogen diffusion induced phase transformation. a,** Schematic illustrations of SCO thin film grown on LSAT (001) (**left**) and DyScO$_3$-DSO (110)$_o$ (**right**) substrates, showing in-plane and out-of-plane ordered oxygen vacancy channels due to the presence of compressive and tensile strains, respectively. O$_O$ and O$_T$ represent the oxygen octahedron and tetrahedron, respectively. **b,** Comparison of XRD $\theta$-$2\theta$ scans for SCO thin films with different orientations of oxygen vacancy, in which the superstructure diffraction peaks disappear once the oxygen vacancy channels turn into the out of plane direction, as observed in the SCO grown on DSO substrate. **c,** Comparison of XRD $\theta$-$2\theta$ scans of SCO

grown on DSO $(110)_o$ substrate before and after the ILG induced protonation, in which the clear shift of the diffraction peaks toward lower diffraction angle indicates the successful protonation of SCO into HSCO. Phase transformation through the Pt catalysis process at 100 °C monitored by *in-situ* XRD with two different proton diffusion pathways, i.e., along vertical (**d**) and lateral (**e**) directions, respectively. For the case of vertical diffusion, SCO can be completely switched into HSCO after about 200 seconds; while for the lateral diffusion, only the SCO layer right under the Pt capping layer can transform into HSCO, while the rest region remains unchanged even after extended forming duration. Thus, comparing these results with those obtained from the in-plane vacancy ordered samples (**Fig. 2b-d** and **Supplementary Fig. 12**), we can conclude that the phase transformation from SCO to HSCO as well as the ionic conductivity in HSCO were strongly dominated by the hydrogen diffusion along the oxygen vacancy channel.